\documentstyle[aaspp4]{article}
\lefthead{McClure}
\righthead{The R Stars}

\begin{document}

\title{The R Stars:  Carbon Stars of a Different Kind}

\author{Robert D. McClure}
\affil{Dominion Astrophysical Observatory,
    Herzberg Institute of Astrophysics,\\
    National Research Council, Canada,
    5071 W. Saanich Road, Victoria, BC, V0S 1N0\\
    Electronic mail:  mcclure@dao.nrc.ca}

\begin{abstract}

   After $\sim$16 years of radial-velocity observations of a sample of
22 R-type carbon stars, no evidence for binary motion has been detected
in any of them.  This is surprising considering that approximately 20\%
of normal late-type giants are spectroscopic binaries, and the fraction
is close to 100\% in barium, CH, and subgiant/dwarf CH and barium
stars.  It is suggested, therefore, that a process that has caused the
mixing of carbon to the surface of these stars cannot act in a wide
binary system.  Possibly, the R stars were once all binaries, but with
separations that would not allow them to evolve completely up the giant
and asymptotic giant branchs without coalescing.  This coalescence may
be the agent which causes carbon produced in the helium-core flash to
be mixed outwards to a region where convection zones can bring it to
the surface of the star.

\end{abstract}

%\keywords{stars: carbon, binary --- stellar evolution}

\section{Introduction}

The Henry Draper classification divided carbon stars into two groups N
and R on the basis of their spectral features.  The N
stars exhibit very strong depression of light in the violet part of the
spectrum.  They are the classical carbon stars that are most easily
detected in infrared surveys and used as tracers of an intermediate age
population in extragalactic objects.  The R stars, on the other hand,
have warmer temperatures, and blue/violet light is accessible to
observation, and atmospheric analysis.

The most extensive analysis of the R stars has been done by Dominy
(1984), who found that in the warm R0$-$R4 stars, C is overabundant on
average by approximately 0.7 dex relative to the sun and to normal G
and K giants.  On the other hand he found that the O abundance is
normally near solar, and the N abundance just slightly enhanced.  At
CNO cycle equilibrium both C and O are depleted substantially.  The
lack of oxygen and carbon depletion, therefore, led Dominiy to the
conclusion that the CNO cycle operating near equilibrium is not
responsible for the fact that C/O $>$ 1 in these stars.  The excess
carbon has to come from some other process.

Dominy also found that the s process element abundances are nearly
solar, similar to results of Gordon (1968), and Green et al. (1973).
This is in sharp contrast with most other carbon and carbon-related
stars (see reviews by McClure 1984a, 1985).  The N, S, barium, CH, sgCH,
and dwarf carbon stars all have significantly enhanced abundances of
the s process elements relative to iron (e.g.,  Lambert 1985; Green and
Margon 1994).  The N stars and many of the S stars are assumed to have
undergone helium-shell flashing on the asymptotic giant branch (AGB),
and the third dredge-up phase has brought carbon and s process elements
to the surface (e.g., see Iben and Renzini 1983).

Like many of the barium, CH, sgCH and dwarf carbon stars, the warmer R
stars are too faint to be on the AGB.  Scalo (1976) has reviewed the
status of various carbon-related stars in the Hertzsprung-Russell
diagram, and has pointed out that the R0$-$R4  stars lie in a
similar temperature/luminosity domain as the barium stars.  They are
well below the lower luminosity boundary of the onset of helium
shell flashing.  The lack of s process enhancement, and low luminosity
led Dominy (1984) to conclude that the R stars probably received their
enhanced carbon from the helium-core flash at the tip of the first
ascent giant branch.  This still remains the most viable hypothesis,
although it is not known how the core-flash contaminants have been
mixed to the surface of the star.

\section{Rationale for Radial Velocity Surveys of Carbon-Related Stars}

In the late 1970's, it had become apparent that the classical carbon
stars could be explained by helium-shell flashing on the AGB, whereas
many of the other stars with enhanced carbon could not, because they
were too faint for AGB evolution.  McClure (1979), suggested the
possibility that the CH stars were binaries, because in globular
clusters they tended to be found only in ones of very low central
concentration where binaries may not be disrupted as easily.  A
velocity survey of the barium stars quickly led to the conclusion that
all barium stars are binaries (McClure, Fletcher and Nemec 1980), and
later surveys showed that CH stars as well as barium stars are all
binaries (McClure 1983, 1984b), presumably having undergone
mass-transfer episodes.  These results were confirmed by observations
of Jorissen and Mayor (1988), who also showed that a large fraction of
S stars are binaries.  The latter stars are found to belong to two
groups (e.g. Brown et. al. 1990 and references therein).  Those that
exhibit the radioactive s process element Tc are single AGB stars
undergoing helium-shell flashing that is slowly raising the C/O ratio
to greater than unity.  Those that do not exhibit Tc are found to be
all binaries that have received material contaminated with C and s
process elements from a companion that had undergone helium-shell
flashing on the AGB.  Thus a binary mass-transfer hypothesis has
explained most of the carbon-related objects that are either too faint
to be on the AGB, or that do not exhibit radioactive Tc, which has a
half life that is a good fraction of a star's lifetime on the AGB.  The
peculiar abundances in these stars all originated from the AGB.

The warmer R stars suffer the same problem in our understanding that we
had in the late 1970's for the barium and CH stars.  They are too faint
to be AGB stars.  Therefore, when it was found that barium stars were
likely all binaries, a survey was begun of radial velocities of the R
stars also.  Note, however, that at that time the s process abundance of
the R stars was unclear.  Therefore, it was considered quite possible
that the R stars would fit into the same AGB $-$ carbon star connection
as the barium and CH stars.

\section{The Stellar Sample, and Radial Velocity Observations}

A sample of 38 R stars (based on their HD classes) was picked,
mostly from the compilation by Yamashita (1972), who examined spectra
of a large number of carbon stars, and identified strengths of various
bands and lines for each star.  Among the R stars, the warmer R0$-$R4
stars are the most similar to normal giants, and less likely to exhibit
velocity ``jitter'' characteristic of more luminous giants which have
more extended and less stable atmospheres.  Mostly R0$-$R4 stars were
included in the sample, therefore, but a few R5-R9 stars were also
observed.  The problems of sorting out other types that may contaminate
the sample (in particular the CH stars) will be discussed below.

The observations were performed using the radial-velocity spectrometer
on the 1.2-m telescope of the Dominion Astrophysical Observatory in
Victoria (see Fletcher et al. 1982; McClure et al. 1985 for
descriptions of the instrument).  Although this instrument is capable
of a precision of about $\pm$ 0.3 km s$^{-1}$, because of the faintness
of the R stars and the limited time spent per observation, the final
errors were larger than this.  From repeated observations, the standard
deviation of a single observation was found to be on the average $\pm
0.62$ km s$^{-1}$.  The radial-velocity values and Julian dates
($-2400000^d$) for all observations in the sample are listed in Table 1.

%\placetable{tbl-1}
%\begin{table}
%\dummytable\label{tbl-1}
%\end{table}

It is very difficult to pick a pure sample of a particular type of
carbon star because they are defined by spectral characteristics which
may be difficult to detect at low dispersion, and these characteristics
sometime overlap between different stellar-population types.  The
approach taken here was to select a fairly large sample of stars that
had been classified as type R, and worry later about sorting out
misclassified objects, when better data and classifications became
available.  Unfortunately, as a result of this, some non-R stars have
been included in my previous preliminary estimates of binary frequency
(e.g., McClure 1989).  The final adopted classifications for stars
in the sample are listed in Table 2.

%\placetable{tbl-2}
%\begin{table}
%\dummytable\label{tbl-2}
%\end{table}

\subsection{The N stars and R5-R9 stars:} \label{nstars}

With the most recent classifications, especially those of
Keenan (1993) and Barnbaum et al. (1996), there are four stars in the
sample that should be rejected because they are N stars.  The velocities for
these four stars are shown at the top of Figure 1.  The scatter in
velocities for each of them is significantly greater than
the precision of the spectrometer, and greater than the scatter for the
true R stars in the sample.  The average of the standard deviations of
the individual observations is $\pm$ 1.62 km s$^{-1}$.  However, there is no
evidence for long-term binary motion, the scatter being just a ``jitter'' which
is typical for high luminosity stars with extended atmospheres (e.g.,
Pryor et al. 1988).

%\placefigure{fig1}

Eight stars in the sample are classified as R5-R9, and the velocity
data for these are shown in the middle part of Figure 1.  These stars
again are well up on the giant branch, and they exhibit a similar
velocity jitter.  The average of the standard deviations for the
individual observations in this case is $\pm$ 1.60 km s$^{-1}$.  Again,
there is no convincing evidence for binary motion.  The star with the
largest deviations, HD 59643, is the coolest star in the sample, with 
a classification of R9.

\subsection{The CH Stars}

The problem of distinguishing between CH stars and R stars is a very
difficult one, especially if the spectra are at rather low resolution
as is the case for many of these stars.  It is very important that this
be done, however, because the CH stars have been shown to be binaries
(McClure 1984b), and since we are trying to determine the binary
frequency for the R stars here, we do not want CH binaries
contaminating the sample.  The CH stars were mostly classified as R
stars before they were recognized as a separate class by Keenan
(1942).  The main distinguishing spectral features are very strong CH
bands, enhanced lines of s process elements and weaker Fe group elements
as well as various strengths of C$_2$ bands.  Many R stars also have
quite strong CH, however (see Yamashita 1972, Figure 10e), and at low
dispersion where the strengths of narrow lines are difficult to estimate,
a CH star may look very similar to an R star.  The CH stars are
mostly high-velocity-halo objects, whereas the R stars belong to the
low-velocity-old-disk population (McLeod 1947).  Therefore, any
high-velocity object can normally be assumed to be a CH star.  There
are, however, a few low velocity CH stars, which Yamashita (1975a) has
identified spectroscopically, and referred to as ``CH-like'' stars.
Their spectra appear to be identical to the CH stars.  If stars
of this type are misclassified as R stars, they will contaminate the
sample, and may affect the estimation of binary frequency.

One star, HD 16115,  which has been classified by Yamashita (1975a), and
by Keenan (1993) as a CH-like star, has been kept in the R star list on
the basis of its s process and Fe abundance.  Although CH bands are strong,
Dominy (1984) finds a nearly normal abundance at high dispersion for both
the s process elements and Fe in this star, and this is one of the main
distinguishing features between a CH and an R classification.

Four other stars, whose velocities are exhibited at the bottom of
Figure 1, have been classified as CH or CH-like.  HD 85066 remained
with an R3 classification in the compilation by Yamashita (1972),
although it exhibits his strongest CH index value.  Hartwick and Cowley
(1985), on the other hand, have classified it more recently at higher
resolution as a CH star.  This star is a definite binary with a period
of 2902 days.  The other three stars have been classified as CH-like by
Yamashita (1975a).  BD~$+2^\circ3336$ is a definite binary with a period of 446
days.  BD~$+29^\circ95$ exhibits erratic velocity behavior, although it cannot
be due to long term binary motion.  Perhaps there has been a difficulty
with identification on some nights at the telescope, but because it is
not an R star, it is not important for the present subject.  One other
CH-like star HD 197604 shows no evidence for binary motion.  If one
neglects BD~$+29^\circ95$, two out of three CH-like stars are binaries, which
is not inconsistent with the high binary frequency of their
high-velocity counterparts.

Orbital elements are listed in Table 3 for the two CH stars that are
definite binaries.  The velocities as a function of phase, resulting
from the orbital solutions are plotted in Figure 2.

%\placefigure{fig2}

%\placetable{tbl-3}
%\begin{table}
%\dummytable\label{tbl-3}
%\end{table}

\subsection{The R0$-$R4 Stars}

The velocities for the remaining 22 R0$-$R4 stars are shown in Figure 3.
The range of dates in the plots is from mid-1979 to mid-1996.

%\placefigure{fig3}

\section{The Binary Frequency of the R Stars}

One would be hard pressed to see any evidence for binary motion from
these R0$-$R4 star data.  This is surprising considering that the binary
frequency for normal G-K giants appears to be of the order of 20\%.

Gunn and Griffin (1979) comment that in their velocity observations of
field giants, about 30\% of stars are variable with amplitudes of a few
km s$^{-1}$ over times scales of a few years.  Mermilliod and Mayor
(1992) have carried out the most extensive systematic velocity program
for binary detection among late-type giants, having surveyed numerous
stars in 177 open clusters.  They find 187 binaries among 905 giants,
or a frequency of $\sim$21\%.  Monitoring of a small random sample of
39 K giants by the present author, the same sample as discussed in
Harris and McClure (1983) supplemented by more recent observations,
shows six binaries, or a frequency of 15\%.

Comparing zero binaries among 22 R stars, with this large combined
sample of normal giant stars, one can carry out a $\chi^2$ test of
homogeneity in a $2\times2$ contingency table (Johnson and
Bhattacharyya 1985).  This gives a $\chi^2$ of 5.621,  indicting that
the possibility that the R stars have a normal binary frequency can be
rejected at the 98\% confidence level!  If one were to retain R5 stars
in the sample, this would provide five more non-binaries, since  even with the
small velocity jitter in these, long term velocity variation can be
ruled out.  They just add to the significance of this result.

\section{The R Stars as Coalesced Binaries?}

It appears very likely that some process that produces carbon in the
atmospheres of R stars cannot operate in a binary system, or
alternatively, causes any original binary companion to disappear.  The
first possibility seems remote.  It is difficult to think of such a
process that will operate in single stars but not in wide binary
systems.  Likely, given the luminosities of the warmer R stars, the
process in question is one that mixes helium-core-flash material to the
surface of the star.  Possibly the R stars were all binaries at one
time, and the process in question caused the disappearance of their
companions.  The most likely event would be the coalescence of a binary
system that had a separation of components that was too small for
evolution up the giant and asymptotic branches to proceed normally.
Coalescence perhaps causes mixing in the vicinity of the core during or
after the helium-core flash.

Smith and Demarque (1980) discuss the effects of mixing at the
core flash in the context of trying to explain the existence of the
sgCH stars.  They showed that the star becomes fainter than the
normal horizontal branch as hydrogen is mixed into the core.  During
the second ascent up the giant branch, convection zones might then reach
down to bring the carbon produced in the flash up to the surface.  They
concluded that the sgCH stars were too faint to be explained in
this manner.  However, the R stars are significantly brighter than
this, and Smith and Demarque's analysis could very well apply to them,
with coalescence of a companion being the process that causes the
mixing. If the R stars are produced in this way, one would expect no
subgiant or dwarf R stars, and this does appear to be the case.  The
subgiant and dwarf carbon-related stars that have been analyzed all
have enhanced s process elements (Green and Margon 1994; Bond, private
communication), unlike the R stars.

%\placefigure{fig4}

One might expect, given the present hypothesis, that the
resulting star should be rotating.  Angular momentum would be gained by
the stellar atmosphere from the orbital energy of the companion
spiraling in to the center of the R star.  For this reason,
observations were made in July 1996 to measure the width of
spectral lines in several R stars, and in normal K giants for
comparison.  Cross-correlation line profiles of these  stars were
observed with the same radial-velocity spectrometer which was used for
the radial-velocity measurements, but care was taken to obtain a
complete profile including continuum on each side of the line.  Figure
4 shows this comparison for two R stars (dots) relative to two K giants
(solid curves).  There is no evidence for rotational broadening  of the
R star profiles, so that this must be taken into account in any binary
evolution explanation.  Angular momentum gained by a model stellar
atmosphere must be dissipated or transferred inwards for this
hypothesis to remain viable.

\acknowledgments

I wish to thank Doug Bond, Murray Fletcher, Les Saddlemyer, and Frank
Younger for help over the years in operation of the telescope and
radial-velocity spectrometer.  I thank Jim Hesser, Chad Hogan, Les
Saddlemyer, and Andy Woodsworth for contributing observations to this
project over the years.  I also thank Chris Tout for very useful
discussions.

\clearpage

\textheight=7.2in
\tiny

\begin{deluxetable}{cccccccccccccc}
\tablenum{1}
\tablewidth{0pt}
\tablecaption{Radial Velocity Data}
\tablehead{
\colhead{JD}             & \colhead{\phs RV}             &
\colhead{}              &
\colhead{JD}             & \colhead{\phs RV}             &
\colhead{}              &
\colhead{JD}             & \colhead{\phs RV}             &
\colhead{}              &
\colhead{JD}             & \colhead{\phs RV}             &
\colhead{}              &
\colhead{JD}             & \colhead{\phs RV} \\
\colhead{\hspace{-.2em}$-2400000$}                 & 
{ km s$^{-1}$}    & \colhead{}                     &
\colhead{\hspace{-.2em}$-2400000$}                 & 
{ km s$^{-1}$}    & \colhead{}                     &
\colhead{\hspace{-.2em}$-2400000$}                 & 
{ km s$^{-1}$}    & \colhead{}                     &
\colhead{\hspace{-.2em}$-2400000$}                 & 
{ km s$^{-1}$}    & \colhead{}                     &
\colhead{\hspace{-.2em}$-2400000$}                 & 
{ km s$^{-1}$}    }

\startdata

%\tablevspace{1ex}

 \multicolumn{2}{c}{\bf HD 1994    }&&  \multicolumn{2}{c}{\bf HD 16115    }&&  \multicolumn{2}{c}{\bf BD+23 601    }&&  \multicolumn{2}{c}{\bf HD 57884    }&&  \multicolumn{2}{c}{\bf HD 63353    }\nl
44876.832 &$        -47.68          $&& 49241.027 &\phs      15.50           && 46714.855 &$        -11.43          $&& 45384.773 &\phs      51.18           && 45367.895 &\phs      16.87           \nl
44935.809 &$        -45.81          $&& 49262.000 &\phs      15.26           && 47031.934 &$        -12.26          $&& 45410.754 &\phs      51.63           && 45384.797 &\phs      18.53           \nl
44967.824 &$        -46.64          $&& 49695.727 &\phs      15.94           && 47086.910 &$        -11.67          $&& 45613.023 &\phs      51.12           && 45410.777 &\phs      20.94           \nl
45207.941 &$        -43.94          $&& 50142.664 &\phs      15.85           && 47196.730 &$        -11.10          $&& 45669.953 &\phs      50.58           && 45452.758 &\phs      16.55           \nl
45384.668 &$        -46.07          $&&           &                          && 47789.930 &$        -11.49          $&& 45712.805 &\phs      49.08           && 45668.062 &\phs      20.20           \nl
45583.988 &$        -46.28          $&&  \multicolumn{2}{c}{\bf HD 19557    }&& 47872.855 &$        -11.10          $&& 46467.781 &\phs      49.56           && 47086.953 &\phs      20.87           \nl
45597.906 &$        -45.51          $&& 44184.957 &\phn$     -8.03          $&& 48950.984 &$        -11.65          $&& 46715.023 &\phs      50.40           && 47260.777 &\phs      18.87           \nl
45669.734 &$        -44.44          $&& 44198.855 &$        -10.10          $&& 49695.844 &$        -11.27          $&& 47872.965 &\phs      50.47           && 47873.000 &\phs      18.66           \nl
45989.758 &$        -43.33          $&& 44331.742 &$        -10.91          $&& 50129.742 &$        -11.43          $&& 50126.816 &\phs      50.38           && 50123.879 &\phs      20.73           \nl
46341.773 &$        -46.44          $&& 44493.855 &$        -13.31          $&&           &                          && 50127.816 &\phs      48.31           && 50128.809 &\phs      21.24           \nl
46657.977 &$        -46.43          $&& 44606.969 &$        -12.90          $&&  \multicolumn{2}{c}{\bf HD 34467    }&& 50142.754 &\phs      49.06           &&           &                          \nl
46990.934 &$        -44.79          $&& 44876.930 &$        -10.96          $&& 45384.754 &\phs      16.59           &&           &                          &&  \multicolumn{2}{c}{\bf HD 76846    }\nl
47086.832 &$        -45.89          $&& 44935.863 &$        -10.83          $&& 45410.770 &\phs      17.75           &&  \multicolumn{2}{c}{\bf HD 58337    }&& 44304.820 &\phs      25.60           \nl
47141.754 &$        -45.86          $&& 45051.711 &$        -11.07          $&& 45598.020 &\phs      17.51           && 45367.879 &\phn$     -1.97          $&& 44331.797 &\phs      25.13           \nl
47763.949 &$        -43.28          $&& 45207.992 &\phn$     -8.64          $&& 45669.984 &\phs      19.33           && 45384.812 &\phn$     -1.42          $&& 44607.910 &\phs      24.89           \nl
47872.797 &$        -45.83          $&& 45283.961 &\phn$     -9.99          $&& 45712.836 &\phs      15.67           && 45410.805 &\phn$     -1.32          $&& 44668.859 &\phs      24.58           \nl
48467.934 &$        -46.38          $&& 45384.730 &\phn$     -9.94          $&& 46467.758 &\phs      16.48           && 45613.000 &\phn$     -1.76          $&& 44683.855 &\phs      25.06           \nl
49261.859 &$        -46.02          $&& 45410.719 &$        -10.56          $&& 46714.887 &\phs      17.11           && 45670.055 &\phn$     -2.05          $&& 44712.723 &\phs      25.01           \nl
49647.754 &$        -46.19          $&& 45597.965 &\phn$     -9.78          $&& 47086.918 &\phs      17.92           && 45712.863 &\phn$     -0.53          $&& 45053.848 &\phs      24.88           \nl
49695.688 &$        -45.01          $&& 45669.961 &$        -11.40          $&& 48950.977 &\phs      18.00           && 46497.812 &\phn$     -0.45          $&& 45284.012 &\phs      24.80           \nl
50129.625 &$        -47.30          $&& 45989.828 &\phn$     -9.83          $&& 49270.055 &\phs      17.13           && 46715.043 &\phn$     -1.96          $&& 45367.898 &\phs      24.82           \nl
          &                          && 46341.816 &\phn$     -9.48          $&&           &                          && 47260.719 &\phn$     -1.49          $&& 45410.852 &\phs      25.04           \nl
 \multicolumn{2}{c}{\bf BD+29 95    }&& 46714.836 &$        -10.72          $&&  \multicolumn{2}{c}{\bf HD 37212    }&& 47873.012 &\phn$     -1.18          $&& 45668.070 &\phs      24.59           \nl
45597.934 &\phs      17.83           && 47005.953 &$        -10.70          $&& 45598.031 &\phs      33.19           && 50123.766 &\phn$     -1.57          $&& 45712.898 &\phs      24.61           \nl
45669.746 &\phs\phn   5.98           && 47086.895 &$        -11.46          $&& 45669.945 &\phs      33.20           && 50142.789 &\phn$     -0.80          $&& 45852.727 &\phs      25.17           \nl
47086.844 &\phs      17.40           && 47196.703 &$        -10.89          $&& 45712.824 &\phs      31.13           &&           &                          && 46110.906 &\phs      24.76           \nl
47763.988 &\phs      18.32           && 47214.715 &$        -12.22          $&& 46467.805 &\phs      32.22           &&  \multicolumn{2}{c}{\bf HD 58364    }&& 46937.758 &\phs      23.52           \nl
47872.785 &\phs      18.68           && 47789.941 &\phn$     -9.41          $&& 46715.008 &\phs      28.42           && 45367.871 &\phn$     -8.85          $&& 47086.961 &\phs      24.25           \nl
49240.840 &$        -12.20          $&& 47872.812 &\phn$     -9.91          $&& 47087.043 &\phs      34.39           && 45384.820 &\phn$     -7.82          $&& 47260.789 &\phs      24.97           \nl
49261.914 &\phs      17.85           && 49262.023 &\phn$     -9.97          $&& 47872.891 &\phs      28.03           && 45410.809 &\phn$     -7.03          $&& 47873.051 &\phs      24.23           \nl
49647.785 &\phs      23.18           && 49695.762 &$        -12.94          $&&           &                          && 45613.008 &\phn$     -9.13          $&& 48951.039 &\phs      24.36           \nl
49695.707 &\phs      23.05           && 50129.688 &\phn$     -6.86          $&&  \multicolumn{2}{c}{\bf BD+33 1194    }&& 45670.062 &\phn$     -7.27          $&& 50128.820 &\phs      24.71           \nl
          &                          &&           &                          && 45598.027 &$        -57.77          $&& 45712.867 &\phn$     -8.36          $&&           &                          \nl
 \multicolumn{2}{c}{\bf BD+21 64    }&&  \multicolumn{2}{c}{\bf HD 19881    }&& 45669.996 &$        -57.59          $&& 46497.812 &\phn$     -8.10          $&&  \multicolumn{2}{c}{\bf HD 77234    }\nl
44876.840 &\phs      13.94           && 45384.742 &\phs      10.49           && 45712.852 &$        -57.21          $&& 46715.047 &\phn$     -8.79          $&& 44304.852 &\phs\phn   4.70           \nl
44935.777 &\phs      14.01           && 45410.734 &\phs      12.37           && 46467.773 &$        -56.56          $&& 47086.938 &$        -10.19          $&& 44331.805 &\phs\phn   4.53           \nl
45207.945 &\phs\phn   9.66           && 45597.973 &\phs      11.19           && 46714.949 &$        -57.93          $&& 47260.723 &\phn$     -7.85          $&& 44382.770 &\phs\phn   5.45           \nl
45384.688 &\phs      12.29           && 45669.969 &\phs      11.64           && 47031.969 &$        -58.40          $&& 47873.023 &\phn$     -8.22          $&& 44607.926 &\phs\phn   6.13           \nl
45597.945 &\phs      12.46           && 45989.840 &\phs      13.10           && 47086.922 &$        -57.86          $&& 50123.785 &\phn$     -9.52          $&& 44668.910 &\phs\phn   7.40           \nl
45669.762 &\phs      12.29           && 46467.695 &\phs      11.71           && 47872.949 &$        -57.85          $&& 50142.805 &\phn$     -9.49          $&& 44683.871 &\phs\phn   5.07           \nl
45989.820 &\phs      11.71           && 46714.848 &\phs      12.77           && 48950.969 &$        -56.55          $&&           &                          && 44712.730 &\phs\phn   6.20           \nl
46341.801 &\phs      12.56           && 47086.898 &\phs      13.56           && 49270.043 &$        -56.11          $&&  \multicolumn{2}{c}{\bf HD 58385    }&& 45284.020 &\phs\phn   2.86           \nl
46704.973 &\phs      12.12           && 47196.715 &\phs      14.27           && 50129.789 &$        -57.78          $&& 45367.883 &\phs      68.89           && 45367.902 &\phs\phn   5.40           \nl
46714.828 &\phs      12.70           && 47789.938 &\phs      11.81           && 50142.727 &$        -56.73          $&& 45384.789 &\phs      68.85           && 45410.855 &\phs\phn   4.07           \nl
46990.969 &\phs      14.39           && 48232.832 &\phs      13.97           &&           &                          && 45410.793 &\phs      69.44           && 45452.777 &\phs\phn   3.36           \nl
47086.879 &\phs      11.07           && 49262.035 &\phs      13.14           &&  \multicolumn{2}{c}{\bf HD 52432    }&& 45668.055 &\phs      67.81           && 45668.078 &\phs\phn   4.43           \nl
47763.980 &\phs      12.47           && 49695.797 &\phs      10.40           && 44607.883 &\phs      24.30           && 45712.812 &\phs      71.04           && 45712.906 &\phs\phn   5.49           \nl
47872.758 &\phs      12.39           && 50129.609 &\phs      14.88           && 44712.695 &\phs      25.47           && 46467.812 &\phs      70.35           && 45752.902 &\phs\phn   4.30           \nl
49240.863 &\phs      12.93           && 50142.688 &\phs      15.42           && 44877.027 &\phs      23.93           && 47087.035 &\phs      70.88           && 45852.730 &\phs\phn   3.83           \nl
49261.938 &\phs      12.35           &&           &                          && 45046.742 &\phs      22.61           && 47260.770 &\phs      68.97           && 46110.914 &\phs\phn   4.22           \nl
49647.801 &\phs      12.73           &&  \multicolumn{2}{c}{\bf HD 286436    }&& 45051.715 &\phs      22.39           && 47872.980 &\phs      70.99           && 46875.777 &\phs\phn   4.08           \nl
49695.715 &\phs      12.98           && 45597.996 &$        -33.98          $&& 45284.004 &\phs      20.31           && 48951.035 &\phs      66.01           && 47086.969 &\phs\phn   2.58           \nl
50129.664 &\phs      12.23           && 45669.812 &$        -34.22          $&& 45384.766 &\phs      22.17           && 50123.812 &\phs      68.57           && 47260.793 &\phs\phn   5.18           \nl
          &                          && 46467.715 &$        -34.75          $&& 45410.742 &\phs      22.89           && 50142.824 &\phs      68.50           && 47873.062 &\phs\phn   5.31           \nl
 \multicolumn{2}{c}{\bf HD 16115    }&& 46714.863 &$        -34.66          $&& 45452.719 &\phs      24.34           &&           &                          && 48951.043 &\phs\phn   5.34           \nl
44876.914 &\phs      15.29           && 47031.910 &$        -35.50          $&& 45613.016 &\phs      23.70           &&  \multicolumn{2}{c}{\bf HD 59643    }&& 50115.875 &\phs\phn   6.05           \nl
44935.762 &\phs      15.96           && 47086.906 &$        -34.71          $&& 45668.047 &\phs      21.28           && 45384.805 &\phs      41.19           && 50128.828 &\phs\phn   4.48           \nl
45207.988 &\phs      14.67           && 47789.926 &$        -33.75          $&& 45669.938 &\phs      21.55           && 45410.812 &\phs      43.02           &&           &                          \nl
45283.922 &\phs      15.26           && 47872.805 &$        -35.07          $&& 45712.801 &\phs      22.10           && 45452.746 &\phs      39.52           &&  \multicolumn{2}{c}{\bf HD 79319    }\nl
45384.617 &\phs      15.54           && 49262.012 &$        -33.37          $&& 46467.777 &\phs      21.78           && 45613.027 &\phs      38.42           && 44304.832 &\phn$     -0.76          $\nl
45597.949 &\phs      14.55           && 49695.824 &$        -34.51          $&& 46715.016 &\phs      23.55           && 45670.074 &\phs      39.22           && 44331.812 &\phn$     -1.22          $\nl
45669.770 &\phs      15.05           && 50112.707 &$        -33.88          $&& 47086.930 &\phs      22.22           && 45712.875 &\phs      41.01           && 44607.953 &\phn$     -3.22          $\nl
45989.852 &\phs      14.73           && 50129.730 &$        -34.90          $&& 47872.961 &\phs      25.26           && 47086.949 &\phs      38.75           && 44668.852 &\phn$     -2.98          $\nl
46341.809 &\phs      13.96           && 50142.703 &$        -33.28          $&& 48951.000 &\phs      23.83           && 47260.781 &\phs      40.76           && 44683.863 &\phn$     -1.44          $\nl
46704.980 &\phs      15.45           &&           &                          && 50129.824 &\phs      24.46           && 47873.039 &\phs      44.75           && 44712.703 &\phn$     -0.19          $\nl
47031.977 &\phs      13.87           &&  \multicolumn{2}{c}{\bf BD+23 601    }&& 50142.738 &\phs      24.82           && 50123.848 &\phs      49.30           && 45053.832 &\phn$     -2.70          $\nl
47086.887 &\phs      14.48           && 45598.012 &$        -11.88          $&&           &                          &&           &                          && 45284.035 &\phn$     -2.82          $\nl
47789.918 &\phs      14.96           && 45669.977 &$        -10.99          $&&           &                          &&           &                          && 45367.922 &\phn$     -3.48          $\nl
\hline
\tablevspace{-2.5ex}
&&&&&&&\nl

 \multicolumn{2}{c}{\bf HD 79319    }&&  \multicolumn{2}{c}{\bf HD 122547    }&&  \multicolumn{2}{c}{\bf BD+23 2998    }&&  \multicolumn{2}{c}{\bf BD+17 3325    }&&  \multicolumn{2}{c}{\bf HD 197604    }\nl
45410.848 &\phn$     -3.04          $&& 45803.926 &$        -27.20          $&& 49560.816 &$        -32.14          $&& 48019.953 &$        -48.12          $&& 46657.863 &\phs      15.73           \nl
45452.766 &\phn$     -3.05          $&& 46234.891 &$        -28.03          $&& 50143.070 &$        -31.13          $&& 48089.812 &$        -48.13          $&& 46970.902 &\phs      16.36           \nl
45668.086 &\phn$     -2.13          $&& 46657.711 &$        -28.60          $&&           &                          && 48467.859 &$        -48.18          $&& 46990.875 &\phs      16.36           \nl
45712.914 &\phn$     -2.58          $&& 46876.000 &$        -28.55          $&&  \multicolumn{2}{c}{\bf HD 156074    }&& 49560.852 &$        -48.49          $&& 47141.637 &\phs      15.68           \nl
45752.906 &\phn$     -2.38          $&& 46970.777 &$        -28.05          $&& 44305.035 &$        -12.94          $&& 50129.062 &$        -47.78          $&& 47763.883 &\phs      17.07           \nl
45852.738 &\phn$     -2.49          $&& 47260.910 &$        -29.00          $&& 44331.957 &$        -12.44          $&&           &                          && 47872.699 &\phs      16.84           \nl
46110.926 &\phn$     -2.69          $&& 48467.746 &$        -27.63          $&& 44382.852 &$        -12.83          $&&  \multicolumn{2}{c}{\bf HD 163838    }&& 48162.734 &\phs      15.46           \nl
46937.773 &\phn$     -4.39          $&& 49560.738 &$        -27.70          $&& 44435.828 &$        -12.19          $&& 45597.770 &$        -41.65          $&& 48467.906 &\phs      15.86           \nl
47087.020 &\phn$     -3.69          $&& 50128.953 &$        -26.78          $&& 44482.738 &$        -12.18          $&& 45667.645 &$        -41.81          $&& 49175.828 &\phs      16.43           \nl
47260.809 &\phn$     -1.47          $&& 50142.957 &$        -26.63          $&& 44493.742 &$        -12.62          $&& 45803.965 &$        -41.42          $&& 49261.754 &\phs      16.26           \nl
47873.098 &\phn$     -3.08          $&&           &                          && 44668.984 &$        -12.06          $&& 45915.918 &$        -42.29          $&& 49560.922 &\phs      15.87           \nl
48951.047 &\phn$     -3.67          $&&  \multicolumn{2}{c}{\bf BD+30 2637    }&& 44712.980 &$        -13.43          $&& 46341.715 &$        -43.34          $&&           &                          \nl
50126.836 &\phn$     -3.13          $&& 45597.691 &$        -95.94          $&& 44781.824 &$        -12.84          $&& 46657.801 &$        -43.57          $&&  \multicolumn{2}{c}{\bf BD+02 4338    }\nl
50127.844 &\phn$     -3.18          $&& 45713.062 &$        -96.71          $&& 44870.641 &$        -12.45          $&& 46937.879 &$        -42.72          $&& 45667.664 &$        -52.48          $\nl
50128.844 &\phn$     -2.44          $&& 45803.930 &$        -96.74          $&& 44907.633 &$        -12.97          $&& 46970.824 &$        -42.26          $&& 45915.941 &$        -53.38          $\nl
          &                          && 46242.820 &$        -96.18          $&& 44935.590 &$        -12.47          $&& 46990.848 &$        -42.91          $&& 46341.734 &$        -53.17          $\nl
 \multicolumn{2}{c}{\bf HD 85066    }&& 46341.660 &$        -96.68          $&& 45171.742 &$        -12.20          $&& 47057.770 &$        -43.72          $&& 46657.871 &$        -52.97          $\nl
44331.828 &$        -16.01          $&& 46657.746 &$        -95.23          $&& 45207.688 &$        -13.43          $&& 47260.996 &$        -42.68          $&& 46970.906 &$        -53.73          $\nl
44683.883 &$        -11.95          $&& 46876.031 &$        -97.51          $&& 45284.645 &$        -13.77          $&& 48019.965 &$        -41.75          $&& 46990.879 &$        -53.47          $\nl
44712.742 &$        -12.39          $&& 46970.770 &$        -96.42          $&& 45452.891 &$        -13.72          $&& 48089.824 &$        -42.89          $&& 47141.645 &$        -53.15          $\nl
45284.031 &\phn$     -7.00          $&& 47031.773 &$        -96.61          $&& 45597.758 &$        -13.50          $&& 48387.973 &$        -41.96          $&& 47763.934 &$        -53.05          $\nl
45367.914 &\phn$     -5.25          $&& 47057.688 &$        -96.52          $&& 45803.945 &$        -13.44          $&& 48467.887 &$        -42.73          $&& 47872.648 &$        -53.96          $\nl
45410.859 &\phn$     -6.41          $&& 47142.074 &$        -96.84          $&& 45915.789 &$        -12.80          $&& 49175.855 &$        -42.19          $&& 48162.742 &$        -53.34          $\nl
45452.789 &\phn$     -6.47          $&& 47260.973 &$        -95.83          $&& 46111.027 &$        -13.37          $&& 49560.875 &$        -43.05          $&& 48467.914 &$        -52.82          $\nl
45668.094 &\phn$     -5.72          $&& 48019.891 &$        -96.36          $&& 46242.879 &$        -13.13          $&& 50129.070 &$        -43.20          $&& 49175.840 &$        -52.97          $\nl
45712.926 &\phn$     -6.78          $&& 48089.773 &$        -96.71          $&& 46341.707 &$        -12.96          $&&           &                          && 49261.766 &$        -52.97          $\nl
45752.898 &\phn$     -6.65          $&& 48387.887 &$        -96.51          $&& 46657.777 &$        -13.33          $&&  \multicolumn{2}{c}{\bf HD 168227    }&& 49560.930 &$        -53.15          $\nl
45852.746 &\phn$     -8.61          $&& 48467.754 &$        -96.18          $&& 46937.828 &$        -13.51          $&& 46242.863 &$        -14.43          $&&           &                          \nl
46937.801 &$        -18.63          $&& 49560.723 &$        -96.51          $&& 46970.805 &$        -12.36          $&& 46341.699 &$        -14.78          $&&  \multicolumn{2}{c}{\bf HD 218851    }\nl
47031.785 &$        -18.21          $&& 50128.969 &$        -96.01          $&& 46990.824 &$        -12.60          $&& 46657.816 &$        -11.40          $&& 45597.828 &$        -42.76          $\nl
47087.012 &$        -17.74          $&& 50143.016 &$        -95.55          $&& 47141.660 &$        -12.61          $&& 46970.859 &$        -10.64          $&& 45667.820 &$        -42.49          $\nl
47142.031 &$        -17.13          $&&           &                          && 48019.941 &$        -12.45          $&& 46990.867 &\phn$     -9.98          $&& 46341.746 &$        -41.94          $\nl
47260.801 &$        -15.73          $&&  \multicolumn{2}{c}{\bf BD+83 442    }&& 48089.816 &$        -13.03          $&& 47261.004 &$        -11.36          $&& 46412.574 &$        -42.62          $\nl
47316.777 &$        -14.73          $&& 45597.754 &$        -16.10          $&& 48387.945 &$        -12.69          $&& 48019.977 &$        -11.79          $&& 46657.887 &$        -41.97          $\nl
48951.066 &$        -12.50          $&& 45667.609 &$        -15.63          $&& 48467.852 &$        -12.58          $&& 48089.797 &$        -10.53          $&& 46970.961 &$        -42.73          $\nl
50115.930 &$        -16.36          $&& 45803.938 &$        -16.02          $&& 49155.844 &$        -12.70          $&& 48162.719 &\phn$     -9.75          $&& 46990.887 &$        -42.72          $\nl
50128.867 &$        -16.08          $&& 46110.965 &$        -15.94          $&& 49560.836 &$        -12.64          $&& 48467.867 &$        -12.26          $&& 47086.977 &$        -43.30          $\nl
50142.910 &$        -15.75          $&& 46341.672 &$        -15.55          $&& 50129.027 &$        -12.78          $&& 49216.809 &$        -11.11          $&& 47763.891 &$        -44.21          $\nl
          &                          && 46657.758 &$        -15.84          $&& 50143.078 &$        -12.63          $&& 49560.891 &$        -12.71          $&& 47872.730 &$        -42.80          $\nl
 \multicolumn{2}{c}{\bf BD+02 2446    }&& 46876.043 &$        -15.74          $&&           &                          &&           &                          && 48162.801 &$        -42.69          $\nl
45712.938 &\phs      10.23           && 46970.785 &$        -15.77          $&&  \multicolumn{2}{c}{\bf BD+02 3336    }&&  \multicolumn{2}{c}{\bf BD+28 3530    }&& 48467.922 &$        -42.43          $\nl
45852.754 &\phs      10.50           && 46990.816 &$        -15.71          $&& 45597.727 &$        -21.19          $&& 45597.797 &\phn$     -4.92          $&& 49175.844 &$        -41.80          $\nl
46937.816 &\phs\phn   9.67           && 47057.754 &$        -16.07          $&& 45803.953 &$        -34.46          $&& 45667.633 &\phn$     -5.19          $&& 49261.789 &$        -41.58          $\nl
46970.754 &\phs\phn   9.61           && 47087.027 &$        -15.92          $&& 46242.840 &$        -36.26          $&& 45915.902 &\phn$     -4.57          $&& 49560.973 &$        -43.09          $\nl
47141.973 &\phs      10.63           && 47260.977 &$        -16.08          $&& 46341.688 &$        -22.87          $&& 46341.723 &\phn$     -5.23          $&&           &                          \nl
47260.812 &\phs      10.62           && 48019.930 &$        -15.65          $&& 46657.785 &$        -38.69          $&& 46657.836 &\phn$     -5.43          $&&  \multicolumn{2}{c}{\bf HD 223392    }\nl
48019.828 &\phs      11.42           && 48089.781 &$        -15.81          $&& 46937.836 &$        -22.85          $&& 46970.867 &\phn$     -4.42          $&& 44485.828 &$        -19.90          $\nl
50123.953 &\phs      11.33           && 48387.957 &$        -13.32          $&& 46970.809 &$        -26.60          $&& 46990.871 &\phn$     -4.45          $&& 44870.805 &$        -19.68          $\nl
50128.902 &\phs      10.47           && 49155.887 &$        -15.45          $&& 46990.836 &$        -29.86          $&& 47057.805 &\phn$     -5.22          $&& 44907.668 &$        -19.69          $\nl
          &                          && 50128.996 &$        -16.37          $&& 47260.984 &$        -19.45          $&& 47763.879 &\phn$     -4.89          $&& 44935.727 &$        -20.12          $\nl
 \multicolumn{2}{c}{\bf BD+04 2735    }&& 50143.039 &$        -15.76          $&& 48019.949 &$        -35.99          $&& 47872.672 &\phn$     -7.73          $&& 45207.902 &$        -20.31          $\nl
45713.039 &$        -15.07          $&&           &                          && 48089.793 &$        -26.74          $&& 48089.859 &\phn$     -4.74          $&& 45597.836 &$        -21.35          $\nl
45803.914 &$        -14.92          $&&  \multicolumn{2}{c}{\bf BD+23 2998    }&& 48467.828 &$        -35.42          $&& 48162.730 &\phn$     -4.87          $&& 45667.684 &$        -20.51          $\nl
45849.824 &$        -13.94          $&& 45597.719 &$        -32.02          $&& 49155.930 &$        -20.61          $&& 48467.895 &\phn$     -4.28          $&& 45915.949 &$        -20.21          $\nl
46242.816 &$        -15.72          $&& 45803.941 &$        -31.87          $&& 49560.844 &$        -17.19          $&& 49175.809 &\phn$     -4.36          $&& 46341.746 &$        -20.75          $\nl
46937.824 &$        -15.68          $&& 45915.797 &$        -30.99          $&& 50143.090 &$        -33.27          $&& 49261.715 &\phn$     -4.71          $&& 46657.930 &$        -20.87          $\nl
46970.742 &$        -14.88          $&& 46242.828 &$        -32.01          $&&           &                          && 49560.906 &\phn$     -5.09          $&& 46990.895 &$        -21.17          $\nl
47142.039 &$        -14.42          $&& 46341.680 &$        -31.85          $&&  \multicolumn{2}{c}{\bf BD+17 3325    }&&           &                          && 47031.832 &$        -21.08          $\nl
47260.906 &$        -14.99          $&& 46657.770 &$        -31.69          $&& 45597.746 &$        -48.28          $&&  \multicolumn{2}{c}{\bf HD 197604    }&& 47141.656 &$        -20.49          $\nl
48019.812 &$        -14.56          $&& 46876.055 &$        -31.98          $&& 45803.957 &$        -48.56          $&& 44482.797 &\phs      16.44           && 47763.938 &$        -20.58          $\nl
48089.770 &$        -14.70          $&& 46970.789 &$        -31.49          $&& 45915.887 &$        -47.10          $&& 44781.855 &\phs      17.00           && 47872.715 &$        -20.99          $\nl
48387.855 &$        -14.79          $&& 47057.727 &$        -31.46          $&& 46242.852 &$        -48.04          $&& 44870.719 &\phs      16.96           && 48162.809 &$        -21.33          $\nl
50128.961 &$        -15.17          $&& 47142.086 &$        -32.33          $&& 46341.691 &$        -47.83          $&& 44907.641 &\phs      16.14           && 48467.926 &$        -20.36          $\nl
50142.941 &$        -15.06          $&& 47316.848 &$        -31.72          $&& 46657.793 &$        -48.28          $&& 44935.605 &\phs      15.64           && 49261.801 &$        -20.36          $\nl
          &                          && 48019.938 &$        -31.59          $&& 46937.844 &$        -48.18          $&& 45207.719 &\phs      16.97           && 49560.980 &$        -20.91          $\nl
 \multicolumn{2}{c}{\bf HD 122547    }&& 48089.785 &$        -31.83          $&& 46970.816 &$        -48.65          $&& 45597.809 &\phs      16.92           &&           &                          \nl
44668.953 &$        -26.52          $&& 48387.930 &$        -31.56          $&& 46990.840 &$        -47.84          $&& 45667.812 &\phs      16.35           &&           &                          \nl
45368.066 &$        -27.45          $&& 48467.809 &$        -31.48          $&& 47057.738 &$        -47.82          $&& 45915.934 &\phs      15.77           &&           &                          \nl
\hline

\enddata
\end{deluxetable}

\footnotesize

\begin{deluxetable}{lllll}
\tablenum{2}
\tablewidth{0pt}
\tablecaption{Carbon Star Sample}
\tablehead{
\colhead{Star} &
\colhead{} &
\colhead{Sp. Class} &
\colhead{} &
\colhead{Ref.}}

\startdata

HD  1994  &\phn\phn\phn\phn & R5, J4.5  &\phn& 1  \nl
BD~$+29^\circ95$   && CH        && 2  \nl
BD~$+21^\circ64$   && R2        && 3  \nl
HD 16115           && R3, CH3   && 3,1\nl
HD 19557           && R5, J4    && 1  \nl
HD 19881           && N5+       && 1  \nl
HD 286436          && R2        && 3  \nl
BD~$+23^\circ601$  && R2        && 3  \nl
HD 34467           && Nb        && 3  \nl
HD 37212           && N4        && 1  \nl
BD~$+33^\circ1194$ && R2        && 3  \nl
HD 52432           && R5, J3.5  && 3,1\nl
HD 57884           && R8        && 3  \nl
HD 58337           && R3        && 3  \nl
HD 58364           && R3        && 3  \nl
HD 58385           && N         && 3  \nl
HD 59643           && R9        && 3  \nl
HD 63353           && R8        && 3  \nl
HD 76846           && R2+       && 1  \nl
HD 77234           && R5        && 3  \nl
HD 79319           && R4        && 3  \nl
HD 85066           && R3, CH    && 3,4\nl
BD~$+2^\circ2446$  && R2        && 3  \nl
BD~$+4^\circ2735$  && R0        && 3  \nl
HD 122547          && R2        && 3  \nl
BD~$+30^\circ2637$ && R0        && 3  \nl
BD~$+83^\circ442$  && R0        && 3  \nl
BD~$+23^\circ2998$ && R2.5      && 1  \nl
HD 156074          && R2        && 1  \nl
BD~$+2^\circ3336$  && CH, N4    && 2,1\nl
BD~$+17^\circ3325$ && R0        && 3  \nl
HD 163838          && R4        && 3  \nl
HD 168227          && R5        && 3  \nl
BD~$+28^\circ3530$ && R0        && 3  \nl
HD 197604          && CH        && 2  \nl
BD~$+2^\circ4338$  && R2        && 3  \nl
HD 218851          && R2        && 3  \nl
HD 223392          && R3        && 3  \nl
\tablerefs{(1) Keenan 1993; Barnbaum et al. 1996 (2) Yamashita 1975a
(3) Shane 1928; Sanford 1944; Vandervort 1958; Yamashita 1972, 1975b;
Stephenson 1973 (4) Hartwick and Cowley 1985}
\enddata
\end{deluxetable}

\normalsize

\begin{deluxetable}{llll}
\tablenum{3}
\tablewidth{0pt}
\tablecaption{Orbital Elements for CH Stars}
\tablehead{
\colhead{Element}             & \colhead{HD 85066}   &
\colhead{}                    &
\colhead{BD~$+02^\circ3336$}}

\startdata
P (days)               & 2901.7$\pm$19.1   && 445.94$\pm$0.61\nl
$\gamma$ (km s$^{-1}$) & $-$13.09$\pm$0.25 && $-$27.46$\pm$0.22\nl
K (km s$^{-1}$)        & 7.40$\pm$0.40     && 10.54$\pm$0.29\nl
e                      & 0.21$\pm$0.04     && 0.03$\pm$0.03\nl
$\omega$ ($^0$)        & 93.9$\pm$8.7      && 162.1$\pm$41.0\nl
T (JD$-2400000^d$)     & 46094.4$\pm$61.8  && 47064.2$\pm$51.0\nl
$a$ sin $i$ (Gm)       & 288.5$\pm$15.8    && 64.6$\pm$1.8\nl
f(m) (M$_\odot$)       & 0.1129$\pm$0.0186 && 0.0541$\pm$0.0045\nl

\enddata
\end{deluxetable}

\begin{figure}
\figurenum{1}
\plotfiddle{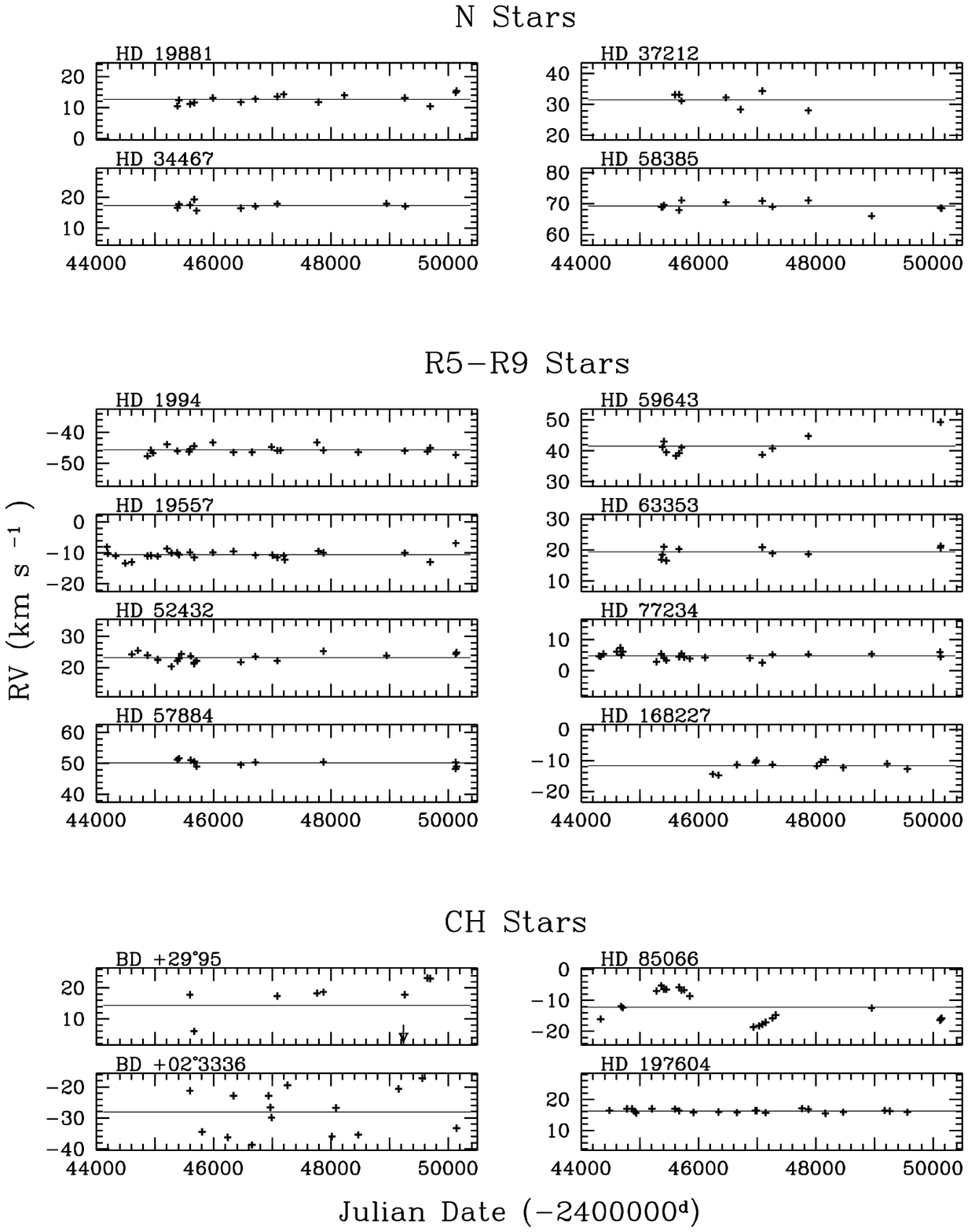}{7.1in}{0degrees}{93}{93}{-290pt}{-90pt}
\caption{Radial Velocities versus Julian Dates for stars in the sample
that have modern classifications of N, R5-R9, or CH-like. \label{fig1}}
\end{figure}

\begin{figure}
\figurenum{2}
\plotfiddle{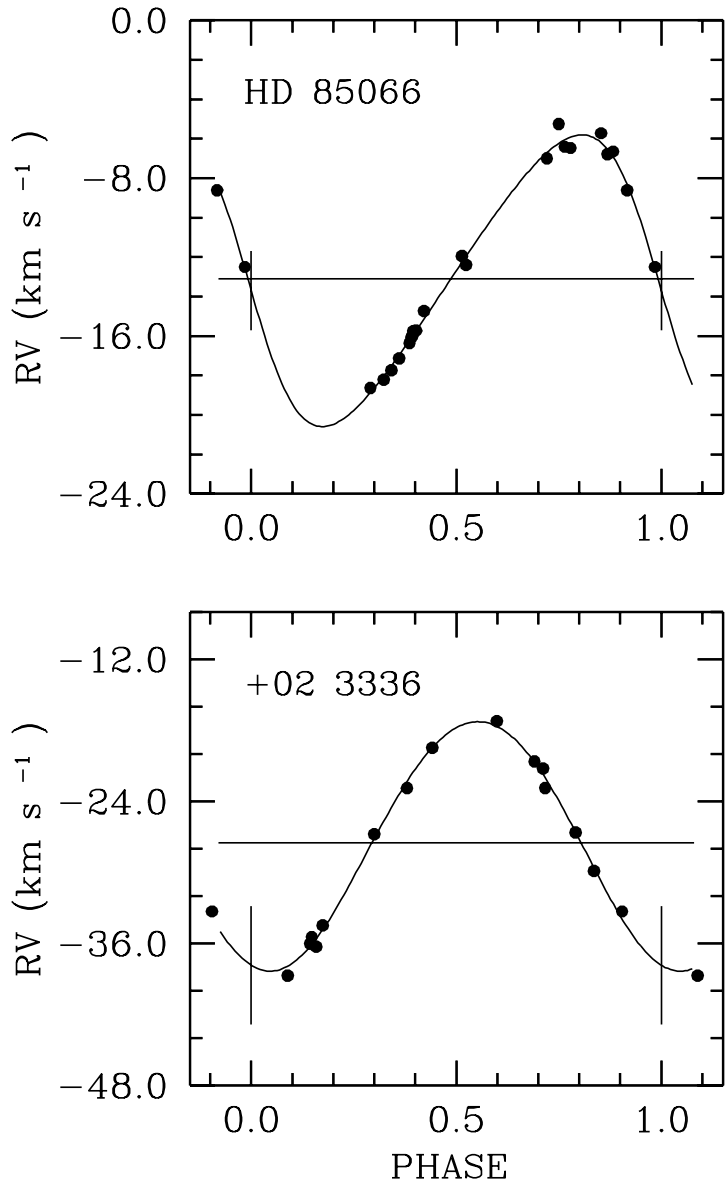}{6.75in}{0degrees}{93}{93}{-290pt}{-110pt}
\caption{ Radial Velocities versus Phase for the computed
orbits (solid curves) from the elements listed in Table 3.  The
observed velocities from which these orbits were calculated are plotted
as dots.  \label{fig2}}
\end{figure}

\begin{figure}
\figurenum{3}
\plotfiddle{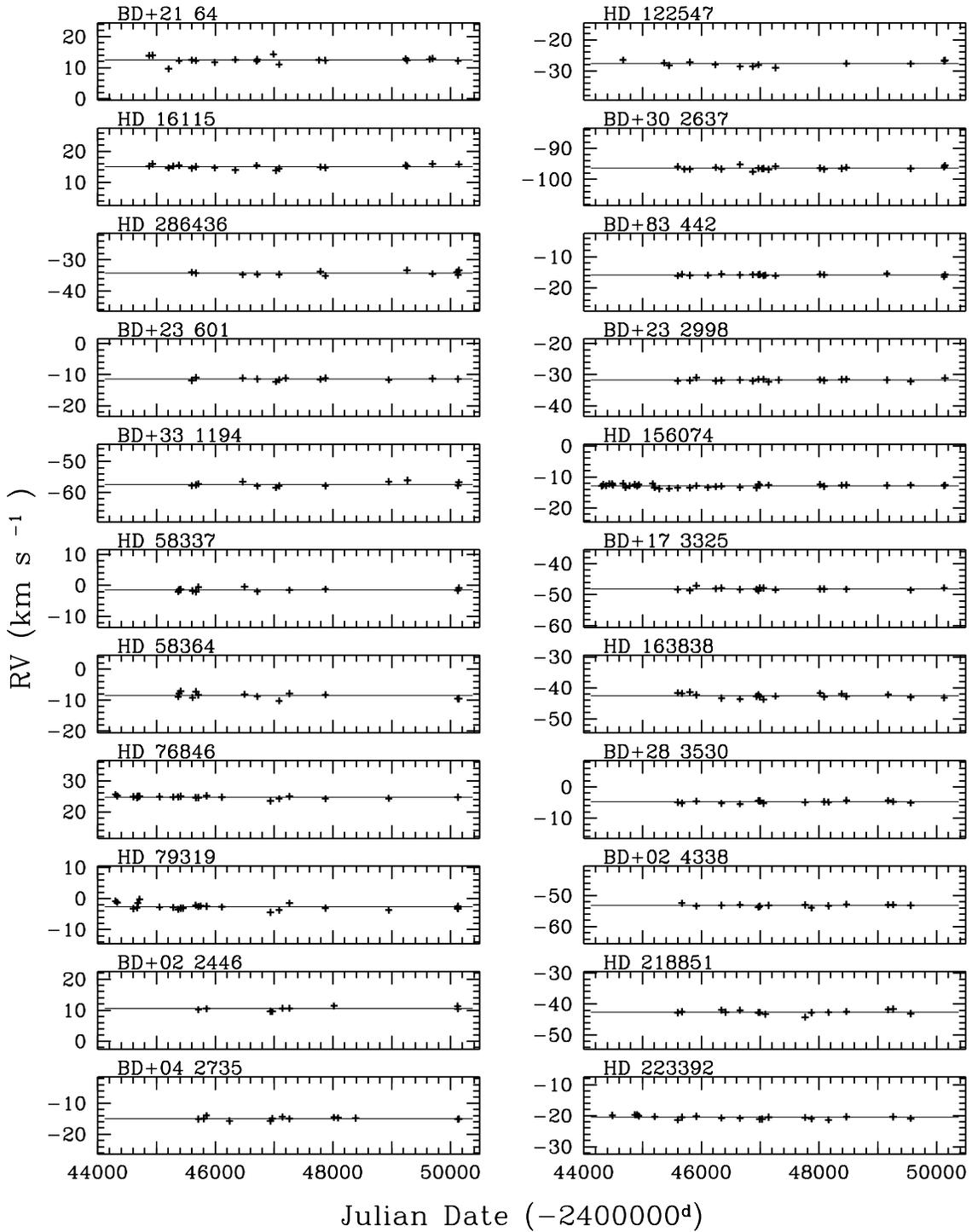}{7.5in}{0degrees}{93}{93}{-290pt}{-90pt}
\caption{Radial Velocities versus Julian Dates for the remaining
R0$-$R4 stars in the sample. \label{fig3}}
\end{figure}

\begin{figure}
\figurenum{4}
\plotfiddle{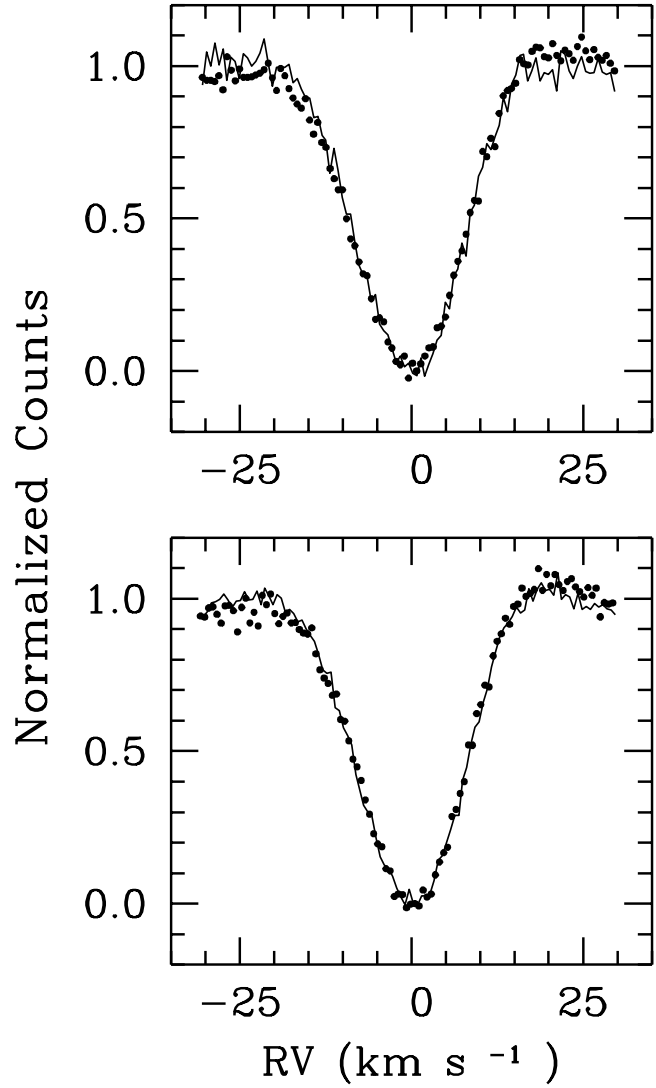}{6.in}{0degrees}{93}{93}{-290pt}{-120pt}
\caption{Comparison of cross-correlation profiles from the
radial-velocity spectrometer for two pairs of R stars (dots) versus
normal K giants (solid curves).  The profiles have been normalized to a
continuum level of 1.0, a minimum of 0.0, and centered on
radial-velocity of 0.0.  The profiles are from the R stars are HD 156074
(left) and BD~$+17^\circ3325$ (right), and the K giants are HD 134493 (left) and
HD 154391 (right). \label{fig4}}
\end{figure}

\end{document}